\documentclass[twocolumn]{aastex631}
\usepackage{mathtools} 
\usepackage{csvsimple}
\usepackage{longtable}
\usepackage{hyperref}

\usepackage{xcolor}

\shorttitle{2024 YR4 and JWST-enabled Planetary Defense Follow-ups}
\shortauthors{de Wit, Rivkin, et al.}

\begin{document}

\title{JWST Observations of Asteroid 2024 YR4 Rule Out a 2032 Lunar Impact and \\ Demonstrate a New Regime for Planetary Defense Follow-up}

\correspondingauthor{J. de Wit, A. S. Rivkin}
\email{jdewit@mit.edu, andy.rivkin@jhuapl.edu}


\author[0000-0003-2415-2191]{Julien de Wit}
\affiliation{Massachusetts Institute of Technology, Cambridge, MA, USA}

\author[0000-0002-9939-9976]{Andrew S. Rivkin}
\affiliation{Johns Hopkins University Applied Physics Laboratory, Laurel, MD, USA}

\author[0000-0001-7895-8209]{Marco Micheli}
\affiliation{European Space Agency, Near Earth Object Coordination Centre, Frascati, Italy}

\author[0000-0003-0774-884X]{Davide Farnocchia}
\affiliation{Jet Propulsion Laboratory, California Institute of Technology, Pasadena, CA, USA}

\author[0000-0001-9892-2406]{Artem Y. Burdanov}
\affiliation{Massachusetts Institute of Technology, Cambridge, MA, USA}

\author[0000-0002-6117-0164]{Bryan Holler}
\affiliation{Space Telescope Science Institute, Baltimore, MD, USA}

\author[0000-0003-0773-1888]{David J. Tholen}
\affiliation{Institute for Astronomy, University of Hawaii at Manoa, Honolulu, HI, USA}

\author[0000-0002-0717-0462]{Thomas Mueller}
\affiliation{Max-Planck-Institut für extraterrestrische Physik, Garching, Germany}

\author[0000-0002-6509-6360]{Maxime Devogele}
\affiliation{European Space Agency, Near Earth Object Coordination Centre, Frascati, Italy}

\author[0000-0003-1582-0581]{Dawn Graninger}
\affiliation{Johns Hopkins University Applied Physics Laboratory, Laurel, MD, USA}

\author[0000-0001-8751-3463]{Heidi B. Hammel}
\affiliation{Association of Universities for Research in Astronomy, Washington, DC, USA}

\author[0000-0001-7694-4129]{Stefanie N. Milam}
\affiliation{NASA Goddard Space Flight Center, Greenbelt, MD, USA}

\author[0000-0003-2467-7713]{Isaac S. Narrett}
\affiliation{Massachusetts Institute of Technology, Cambridge, MA, USA}

\author[0000-0001-8434-9776]{Petr Pravec}
\affiliation{Ondřejov Observatory, Ondřejov, Czechia}

\author[0000-0003-3091-5757]{Cristina A. Thomas}
\affiliation{Northern Arizona University, Flagstaff, AZ, USA}

\begin{abstract}
At the end of its discovery apparition, the $\sim$60 m near-Earth object 2024 YR4 was associated with a non-zero probability of lunar impact during its 2032 December 22 close approach. While posing no threat to Earth, a lunar impact of this scale could have consequences for Earth-orbiting infrastructure, as well as for human exploration on and around the Moon.
We present new JWST/NIRCam observations from 2026 February 18 and 26 that extend the observational arc by eight months, reduce the uncertainty in the 2032 lunar encounter by a factor $>$30, and constitute the faintest detection of a near-Earth object to date, reaching $V \sim 30.5$---beyond the $V \sim 27$ ground-based limit. The updated orbit solution yields a predicted miss distance of $22{\,}900 \pm 800$ km (1$\sigma$) from the center of the Moon, thus ruling out a lunar impact.
Despite challenges due to the limited number of reference stars and saturation and trailing effects, we derive astrometric positions with three independent analysis methods, demonstrating consistency at the $\lesssim$50 mas level. These observations extend the orbital arc at epochs when the object is not accessible from the ground, advancing the timeline for hazard assessment by two years relative to the next feasible ground-based recovery.
This capability is critical in an emerging regime of planetary defense characterized by the discovery of decameter-scale objects by next-generation surveys. These objects are far more common but rapidly become inaccessible to ground-based follow-up. In this regime, hazard assessment can become follow-up--limited, requiring targeted space-based observations, such as those demonstrated here, to reliably constrain impact probabilities on operationally relevant timescales.
\end{abstract}



\section{Introduction}
Planetary defense has historically prioritized the identification and mitigation of large near-Earth objects (NEOs), capable of producing the most severe damage in case of an impact. Over the past decades, survey efforts have substantially improved our knowledge of the risk posed by kilometer-scale NEOs \citep{Harris2015}, shifting attention toward smaller bodies that are more numerous and thus more frequently encountered. While such objects do not pose global hazards, the 2013 Chelyabinsk airburst and the more energetic 1908 Tunguska event demonstrate that  impactors of $\sim$10--100 m in size can produce substantial regional effects \citep{Chyba1993,Brown2013}.

This transition is occurring in parallel with a rapid expansion of observational capabilities. Upcoming facilities such as the Vera C. Rubin Observatory and the space-based missions NASA's NEO Surveyor and Roman Space Telescope, as well as ESA's proposed NEOMIR mission, are expected to transform sensitivity to faint, small bodies, leading to a substantial rise in the discovery rate of decameter-scale NEOs \citep{Ivezic2019, Mainzer2023, Holler2025, Conversi2024}. At the same time, the Earth--Moon system is becoming increasingly populated with space-based infrastructure, introducing new operational considerations for impacts that, while not globally catastrophic, may nevertheless produce energetic events within near-Earth space \citep{Wiegert2025, He2026}.

A central challenge in this emerging regime is the ability to rapidly refine the trajectories of faint objects following discovery. Limited observational arcs, combined with rapid fading below ground-based detection thresholds, often result in extended periods of orbital uncertainty, delaying the conclusive assessment of close approaches or potential impacts \citep{Farnocchia2015}. As survey capabilities expand toward fainter and more numerous objects, this challenge is expected to place increasing demands on follow-up resources and observational cadence, effectively limiting the ability to extend observational arcs for a growing fraction of new objects.

Asteroid 2024 YR4 provides a concrete illustration of this regime, although its observing geometry was relatively favorable compared to typical cases. Following its discovery and early characterization, its orbit was initially consistent with an Earth impact during its 2032 December 22 close approach \citep{Micheli2026, Farnocchia2026,Devog2026}. The last ground-based observation was collected in March 2025, and shortly after the object became too faint for further ground-based tracking until 2028.
By this point, the possible Earth impact was ruled out but a lunar impact was still possible.

Space-based observatories offer a pathway to complement and enhance the information from ground-based data.
Recent work has demonstrated that the James Webb Space Telescope (JWST) can detect and characterize decameter asteroids at distances extending to the main belt, providing size measurements that are largely independent of albedo uncertainties \citep{Burdanov2025}. This capability enables the identification and physical characterization of objects that would otherwise remain beyond the reach of ground-based facilities.

In this context, JWST also provides a unique opportunity to extend observational arcs and refine trajectories for faint objects on timescales directly relevant for hazard assessment. JWST observations of 2024 YR4 have already enabled a measurement of its size \citep{Rivkin2025}.
Moreover, in May 2025, JWST obtained additional astrometry of 2024 YR4, which reduced orbital uncertainties by 20\% and led to a 4\% probability of lunar impact \citep{Farnocchia2026}.
The observations presented here extend this capability by providing high-precision astrometry to constrain its orbit and impact probability well in advance of the next ground-based observing opportunity, illustrating the combined use of JWST for both physical characterization and trajectory refinement of decameter-scale objects.

\begin{figure*}[ht!]
\centering
\includegraphics[width=0.90\textwidth]{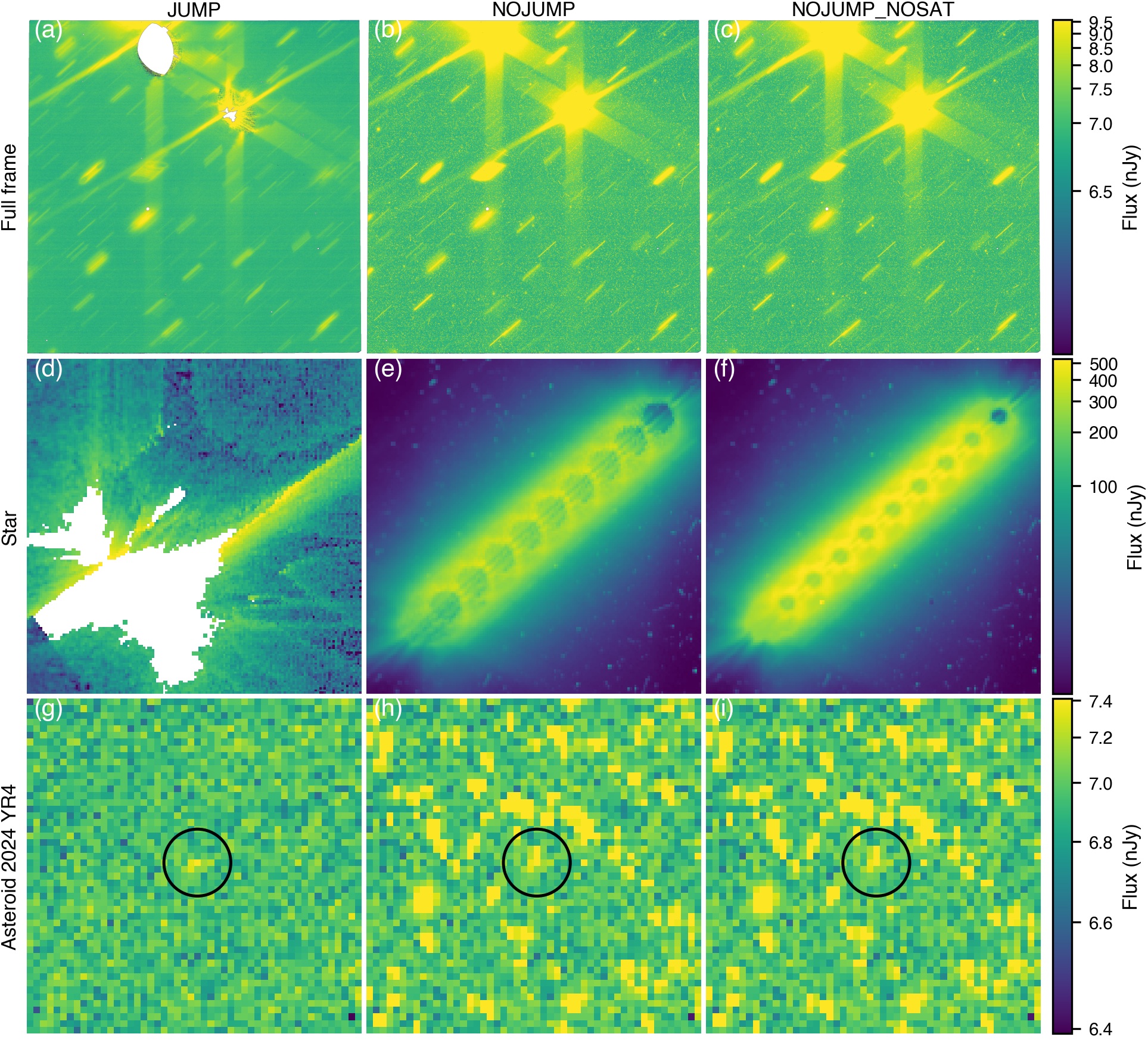}
\caption{\textbf{Impact of NIRCam calibration pipeline settings on star trails and faint-source recovery.}
Three calibrated data products are compared to optimize astrometry for the faint asteroid 2024~YR4. Each column shows (top) a full-frame image from detector NRCB1, (middle) a $300\times300$ pixel cutout centered on a bright stellar trail (Gaia~6284428518676861568; $G=13.8$), and (bottom) a $50\times50$ pixel cutout centered on the asteroid (black circle).
Left: jump step enabled; the asteroid is clearly detected, but saturated stellar cores are removed (white pixels), hindering accurate centroiding of star trails.  
Middle: jump step disabled; stellar cores are better preserved, but asteroid detection is degraded by unflagged cosmic rays.  
Right: jump and saturation steps disabled; stellar trails are best preserved and used for centroiding.
Jump-enabled products are used for asteroid centroiding and photometry, while products with both steps disabled are used for stellar astrometry and plate solving.}
\label{fig:reduction_products}
\end{figure*}

\begin{figure*}[ht!]

\includegraphics[width=0.98\textwidth]{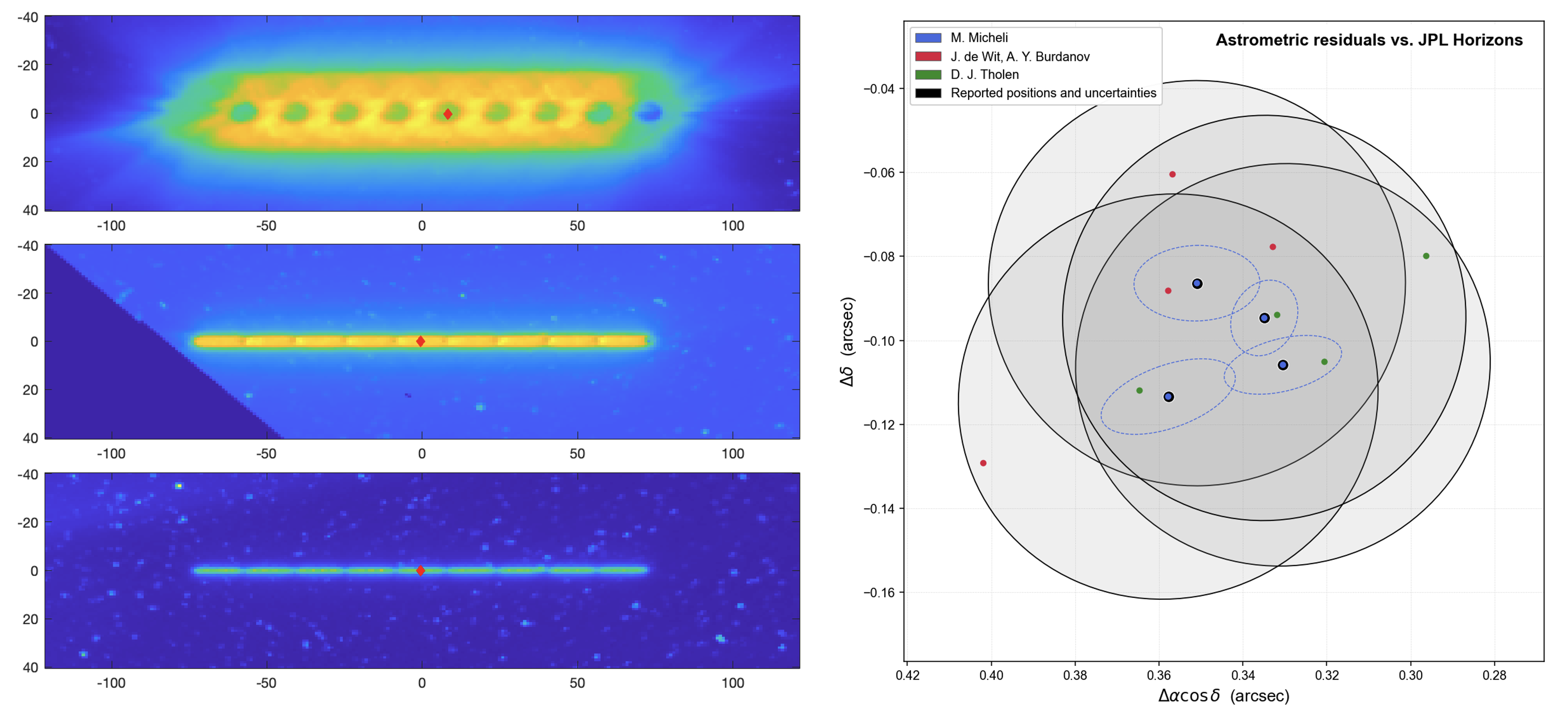}
\caption{\textbf{Astrometric solution using streaked stars and associated uncertainties for asteroid 2024 YR4.}
\textbf{Left:} Representative extraction of the astrometric position of streaked reference stars with different level of saturation in the JWST/NIRCam images. The segmented structure produced by the 9 groups within each integration leaves a visible pattern along the trail, providing additional information on the centroid (red diamond) and trail geometry. This structure helps refine the position measurement of reference stars despite trailing as well as mild (middle panel) and high (top panel) saturation, and is exploited in the plate-solving procedure (see Section~\ref{sec:astrometry}). 
\textbf{Right:} Astrometric residuals of 2024 YR4 relative to the pre-existing JPL Horizons orbit solution (\#78) for the four exposures obtained on 2026 February 26, as derived by three independent pipelines. The colored points show the solutions obtained by the different teams, while the black ellipses indicate the reported positional uncertainties. The close agreement between the independent solutions, well within the adopted \(\sim 50\) mas uncertainty, demonstrates the robustness of the derived astrometry and supports the subsequent orbital refinement.}
\label{fig:astrometry}
\end{figure*}

\section{Observations and Data Reduction}
\label{sec:observations}

JWST observed 2024 YR4 using its Near Infrared Camera (NIRCam) as part of the Director’s Discretionary Time (DDT) program 9441 (PI: Rivkin) on 2026 February 18 and 26 UT. These two visits correspond to the only observing windows in 2026 that simultaneously provide a sufficient number of Gaia reference stars within a single short-wavelength NIRCam detector field of view for high-precision astrometric plate solving and adequate target brightness for robust detection. Together, they therefore enable the most precise orbital constraints practically achievable with JWST during the 2026 apparition, with a targeted positional accuracy of $\sim$100 mas.

The observations used the F150W2 and F322W2 filters, the FULL B module (four short-wavelength detectors and one long-wavelength detector), the SHALLOW2 readout pattern, 10 groups per integration, 9 integrations per dither, totaling $\sim$75\,min of effective exposure time, and a 4-point INTRAMODULEBOX dither. The pointing was offset such that 2024~YR4 fell near the center of the B1 short-wavelength detector.

We retrieved the raw files (uncal) from the Mikulski Archive for Space Telescopes (MAST) and processed with version 1.20.2 of the JWST calibration pipeline \citep{Bushouse2025} and reference file context jwst$\_$1478.pmap. We then generated several types of data products to identify the best method for retrieving astrometry for the extremely faint target (see Fig.~\ref{fig:reduction_products}):
\begin{itemize}
\item {\it Default processing} or \textit{JUMP.} No changes to processing compared to what is available on MAST except for an updated reference file context and the ``clean flicker noise'' step turned on.
\item {\it NOJUMP.} Similar to {\it JUMP}, but with the ``jump'' step turned off. This step flags pixels with changes in flux between groups above a certain threshold and is designed to remove cosmic rays. However, for moving target observations where the target is being tracked non-sidereally, this step removes the cores of the star streaks, which complicate attempts at astrometry. Turning this step off recovers the majority of star streak cores, but also results in the presence of cosmic ray signatures throughout the image.
\item {\it NOJUMP$\_$NOSAT.} 
Similar to {\it NOJUMP}, but with the ``saturation'' step turned off. Even with the ``jump'' step disabled, brighter stars still appear with truncated cores due to saturation in the first two groups. Disabling the ``saturation'' step, which flags and removes saturated pixels during ramp fitting, restores the full shape of the stellar streaks. This configuration yields accurate astrometry for the reference stars, despite unreliable photometry.
\end{itemize}
We generated all products using the ``clean flicker noise'' step that removes 1/$f$ pattern noise. We tested all products and found that for this work images with both the ``jump'' and ``saturation'' steps turned off, which ensured the cores of partially saturated star trails were not removed, were the optimal products for astrometric analysis. 
For the asteroid inferences (photometry and astrometry of the target), we used default processing with the ``clean flicker noise'' step.



\section{Astrometry}
\label{sec:astrometry}

Three subsets of authors performed independent astrometric reductions and analyses to test the sensitivity of our solution to data products, reference-star astrometry, plate-solving strategies, centroiding, and treatment of trailing effects. Despite substantial methodological differences, all solutions converge to consistent astrometric positions at the $\lesssim$50 mas level across all eight JWST/NIRCam exposures (Fig.\,\ref{fig:astrometry}, right panel), demonstrating that the inferred orbit is robust to the choice of reduction strategy. This level of agreement provides strong validation of the astrometric solution given the unusual observational regime of a faint, non-sidereal target embedded in streaked stellar backgrounds. 

We describe here one representative reduction (associated with the solution labeled ``M. Micheli'' in Fig.\,\ref{fig:astrometry}), while detailed descriptions of all individual pipelines are provided in the Appendix, including a discussion specific to the astrometric timetags. Astrometric measurements of 2024 YR4 were obtained using a hybrid approach combining two JWST pipeline data products. The centroid of the target was measured on standard level 2b \texttt{i2d} images (fully calibrated and rectified exposure-level products from MAST) by fitting a radially symmetric Gaussian to the point-spread function. Centroiding uncertainties were estimated from the peak S/N and fitted FWHM. These products are optimal for target extraction, as they are largely free of cosmic rays and artifacts, but substantially alter stellar trails.

The astrometric calibration was instead derived independently using custom-processed images in which jump detection and saturation correction were disabled (i.e., {\it NOJUMP$\_$NOSAT.}), preserving the morphology of stellar trails. Reference stars from the Gaia DR3 catalog \citep{Gaia2023} were identified and their trail centroids measured and compared to WCS-predicted positions. Stars affected by saturation, lacking proper motion information, or truncated by detector edges were excluded. The residual offsets were averaged to derive a global astrometric correction, modeled as a translation (with rotation consistent with zero within uncertainties). This correction was applied uniformly across the detector, consistent with the low level of residual geometric distortion in NIRCam imaging.

For each exposure, this procedure yields a 2-D astrometric correction with associated uncertainties derived from the dispersion of individual stellar offsets. The final position of the target is obtained by applying this correction to the measured centroid, with total uncertainties computed as the convolution of the calibration uncertainty ellipse and the centroiding uncertainty.

While the same general approach was applied at both epochs, differences in observing geometry affected the implementation. During the first epoch, the limited number of usable reference stars (just 2 for some exposures) required a combined solution across detectors, and the relatively slow apparent motion of the target enabled fitting of a continuous trail across the full exposure. During the second epoch, a sufficient number of reference stars (up to 8) allowed independent chip-level solutions, while the faster apparent motion produced well-separated trails for individual integrations, enabling more precise trail fitting over shorter segments.

The two other independent pipelines adopt different implementations for centroiding, data products, and trail modeling, yet yield consistent astrometric positions within the reported uncertainties (see Appendix and Fig.\,\ref{fig:astrometry}). Across all approaches, heavily saturated stars must be excluded from the solution, and the low level of residual distortion ($\sim$1 mas) supports the use of first-order plate solutions based on a small number of reference stars. In all cases, the dominant uncertainties arise from the astrometric calibration rather than centroiding of the target itself.

These February 2026 JWST/NIRCam observations achieved an astrometric precision of $\sim$50 mas on each of the eight exposures\footnote{https://minorplanetcenter.net/mpec/K26/K26E25.xml}.

The results achieved by applying a non-rotational shift determined across all four chips open up some very interesting possibilities for NIRCam's astrometric usability. Because a simple shift is now demonstrated to be sufficient, and can be determined with fewer reference stars (potentially just one, although more are preferred for statistical robustness and uncertainty estimation), scheduling will not need to be restricted to specific observational windows that require a high density of Gaia stars within a single detector.

\begin{figure}[htb]
\hspace{-5mm}
\includegraphics[width=1.1\columnwidth]{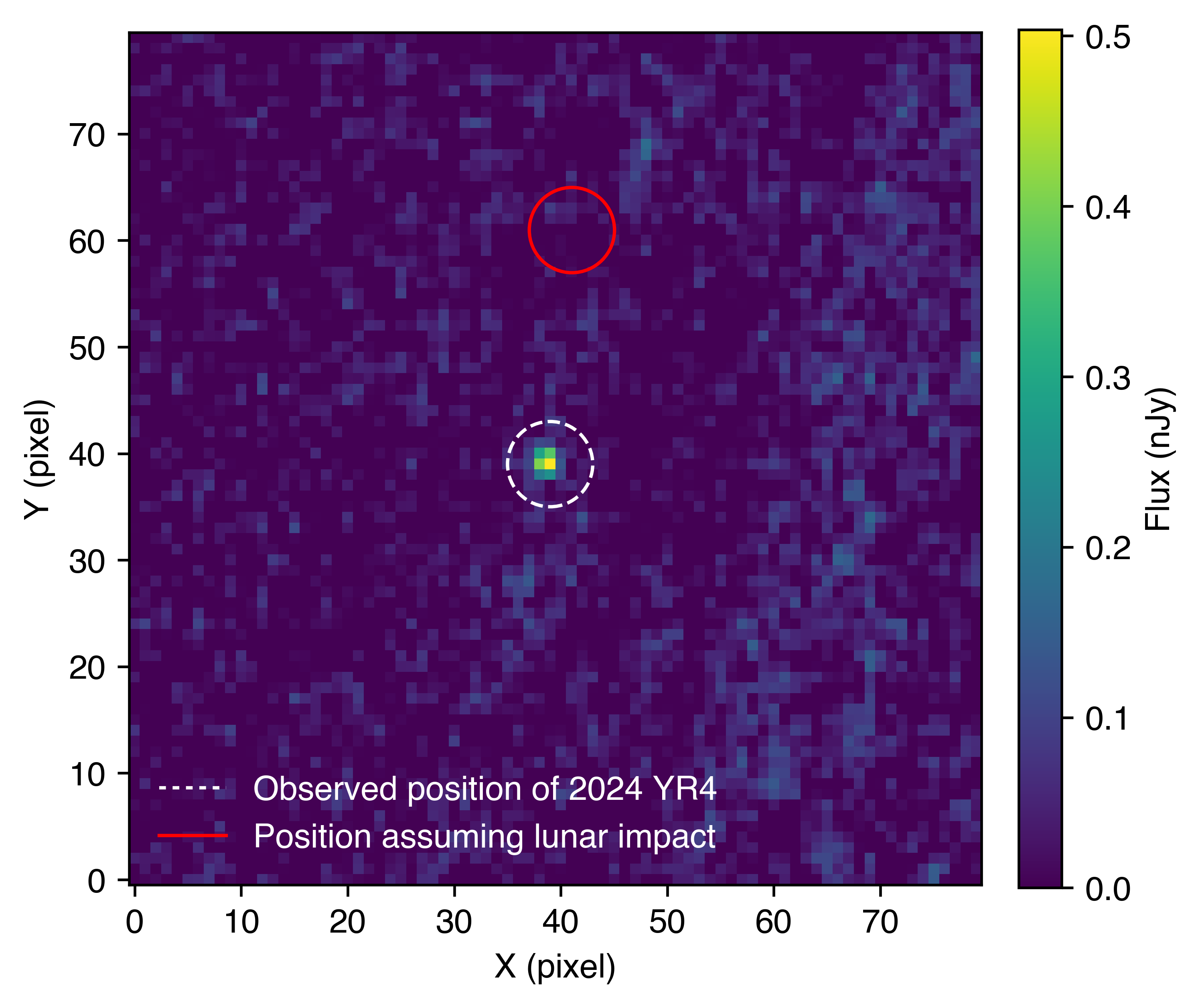}
\caption{\textbf{Plane-of-sky detection of 2024~YR4 at $\mathrm{V_{mag} \sim 30.5}$ and comparison to the position corresponding to a lunar-impact solution.}
Stacked JWST/NIRCam exposures obtained on 2026 February 26 show the high-significance detection of asteroid 2024 YR4 (white dashed circle)  offset by \(\sim 22\) pixels from the position that would have supported a non-zero 2032 lunar-impact probability (red circle).
The detected source has a flux density of only \(\sim4.0\) nJy, corresponding to an approximate visual magnitude of \(V \sim 30.5\), illustrating the extreme faintness regime probed by these observations.}
\label{fig:yr4_detection}
\end{figure}

\section{Photometry}
\label{sec:photometry}

To obtain fluxes of 2024~YR4, we used two Level~3 NIRCam images in the short wavelength channel (F150W2 filter) after the default pipeline processing with jump step turned on. The first image corresponds to the observation conducted on 2026 February 18 (OBS1) and the second one to 2026 February 26 (OBS2, see Fig.\,\ref{fig:yr4_detection}). Each Level 3 image is generated by combining the dithered exposures using the standard drizzle algorithm, including outlier rejection and background matching, and aligned in the moving-target reference frame to optimize the detection of the asteroid. 

We performed aperture photometry using Photutils  \citep{larry_bradley_2025_17129028}. To determine the source centroid, we convolved a 15x15 pixel cutout centered on the predicted position with a Gaussian kernel to identify the peak, and then refined it with a flux-weighted centroid calculation within a 3-pixel radius around the peak. 

We identified the optimal aperture radius by constructing a growth curve that measures the background-subtracted flux through circular apertures with radii ranging from 0.1 to 15.0 pixels with 0.1 pixel step. Each measurement was corrected for the encircled energy fraction using values interpolated from the NIRCam aperture correction reference file. The optimal aperture radius was selected as the one maximizing the signal-to-noise ratio (SNR). The final flux density is the mean of the aperture-corrected measurements within a $\pm$0.5 pixel window centered on the optimal radius. The total uncertainty was computed as the quadrature sum of the statistical uncertainty, estimated from the standard deviation of flux values within the averaging window, and a 5\% systematic uncertainty floor accounting for residual flat-field errors and aperture correction uncertainties. After converting native image surface brightness units (MJy/sr) to flux density,  we measured a flux density of 4.1$\pm$0.6\,nJy for OBS1 and 4.3$\pm$0.6~nJy for OBS2. The flux density was then converted to equivalent $V$ magnitudes using standard flux conversions and assuming solar colors for 2024~YR4, yielding $V = 30.5 \pm 0.3$; the uncertainty introduced by this assumption is small compared to the overall photometric uncertainty.

We did not detect the asteroid in the NIRCam long-wavelength channel (F322W2), allowing us to place a 3-sigma upper limit of 2.5\,nJy on the flux of the asteroid in this band.

\section{Orbit Determination}
\label{sec:orbit}

Bases on the orbit computed at the end of the discovery apparition, JPL solution 78, the 2024 YR4 ephemeris uncertainty in the JWST plane of sky in February 2026 was 0.7$''$.
Therefore, astrometric measurements with an uncertainty $\sim 50$ mas greatly reduce the orbital uncertainties of 2024 YR4.
The asteroid was found 0.5-$\sigma$ away from the prediction, thus confirming the stability and reliability of the orbital solution as already noticed in 2025 \citep{Farnocchia2026}.

We computed an updated orbit solution (\#79) for 2024 YR4 with a least squares fit \citep{Farnocchia2015} to the entire observation dataset from December 2024 to the eight astrometric positions from JWST in February 2026.
These latter positions have small residuals with an RMS of 6 mas, well within their astrometric uncertainties.
Thanks to the eight-month arc extension, the circumstances of the close approach with the Moon are much better known: 2024 YR4 will fly by the Moon at a distance of 22\,900 $\pm$ 800 km at 2032-12-22 14:57:26 TDB $\pm$ 50 s (1$\sigma$ uncertainties).
As a result, the potential impact with the Moon is completely ruled out.

\begin{figure}[htb]
\centering
\includegraphics[width=\columnwidth]{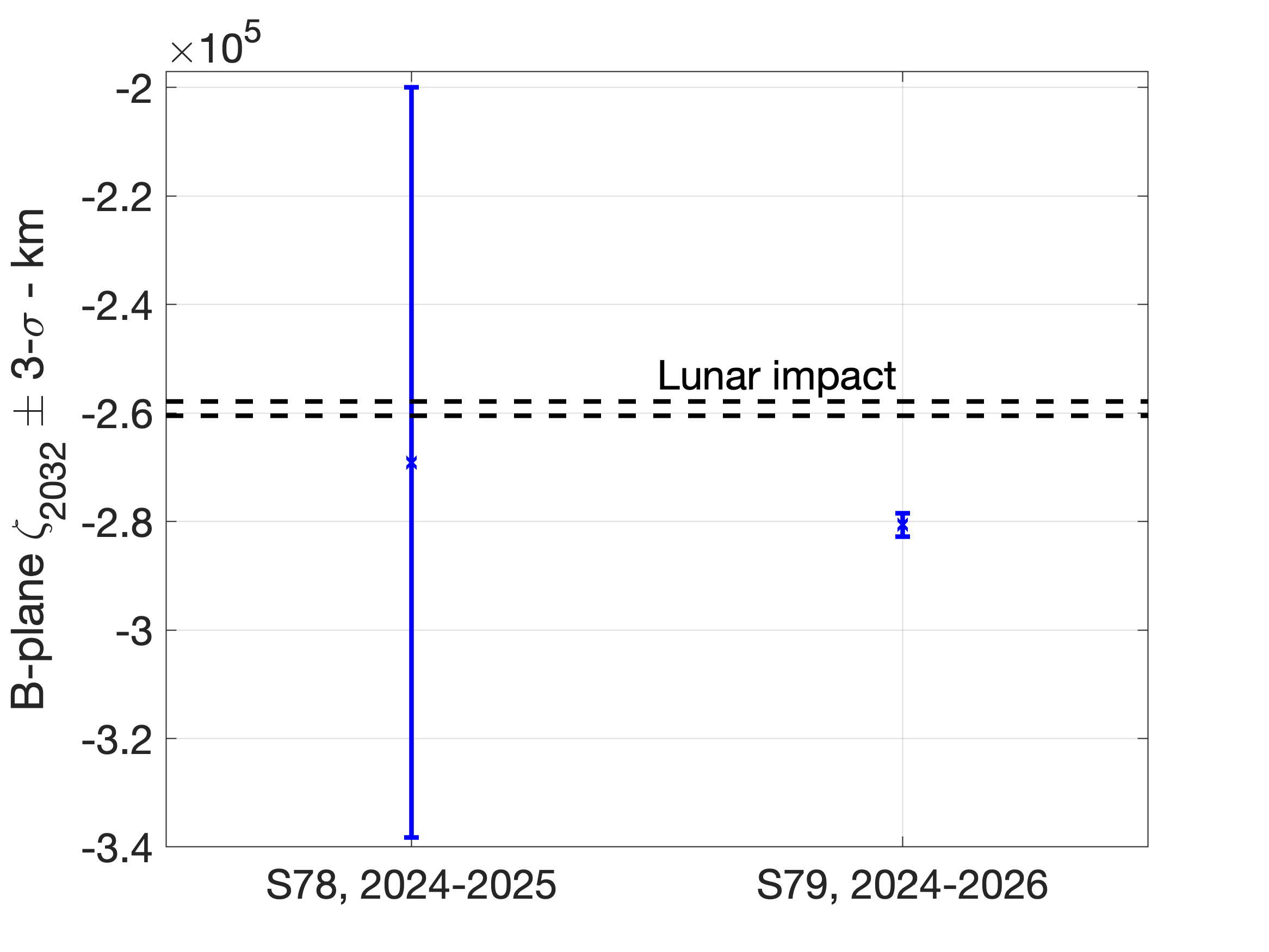}
\caption{\textbf{Prediction for the $\zeta$ coordinate on the Earth B-plane for the 2032 close approach} before (\#78) and after (\#79) including the February 2026 astrometry from JWST. The dashed lines delimit the range of $\zeta$ values corresponding to a lunar impact.}
\label{fig:yr4_orbit}
\end{figure}

Fig.~\ref{fig:yr4_orbit} shows the prediction for the $\zeta$ coordinate on the Earth B-plane \citep{Farnocchia2019} for the 2032 close approach.
For orbit solution \#78, which was the final one from the discovery apparition, $\zeta_{2032} = -269\,000 \pm 23\,000$ km, while for the updated solution \#79 $\zeta_{2032} = -280\,610 \pm 720$ km.
Therefore, the prediction improved by a factor of 32 and the new solution is incompatible with a lunar impact, which corresponds to $-260\,500 < \zeta_{2032} < -257\,870$ km.

Given the small size of 2024 YR4 and the arc extension, the Yarkovsky effect \citep{Vokrouhlicky2015} becomes a possible consideration.
Although there is no signal for a Yarkovsky drift in the dataset, it is possible that this nongravitational perturbation increases prediction uncertainties.
Given a diameter of 60 m, we computed an a priori range for the Yarkovsky parameter $A_2$ of $0 \pm 5\times10^{-13}$ au/d$^2$ \citep{Farnocchia2013}.
The resulting dispersion yields an inconsequential 5\% increase in the $\zeta_{2032}$ uncertainty.
Solar radiation pressure \citep{Vorkouhlicky2000} has even a smaller effect on the trajectory of 2024 YR4.
Therefore, a gravity-only modeling is sufficiently accurate to predict the 2032 encunter at the current level of uncertainty.

Finally, we analyzed the possibility of Earth impacts past the 2032 encounter using Sentry \citep{Roa2021}.
The previously possible low-probability impact solution for 2047 is now ruled out, as well as any other Earth impact over the next 100 years.

\section{Discussion}
\label{sec:discussion}

JWST's observations of 2024 YR4 demonstrate the telescope's potential in planetary defense.
Already in 2025, JWST not only provided an estimate of the asteroid's size \citep{Rivkin2025} but also extended the data arc of 2024 YR4 one month past what was possible from the ground \citep{Micheli2026}, resulting in a 20\% improvement in the prediction of the 2032 close approach to Earth.
The February 2026 observations discussed in this paper are even more compelling: they represent the faintest detection of a near-Earth asteroid and reduced orbital uncertainties by a factor of $>30$, enabling a conclusive assessment of the potential lunar impact in 2032---two years prior to the next observing opportunity from the ground.

JWST's capability of detecting small bodies $\sim$3--4 magnitudes fainter than from the ground represents a new regime for planetary defense follow-up of potential impactors.
This is especially relevant for smaller objects, whose observing windows are short and for which extending the observational arc with JWST may become critical.

Ground-based follow-up of such objects is inherently limited to short observational arcs, as they rapidly fade below detection limits following discovery. In this regime, orbit solutions can retain substantial uncertainties at subsequent close approaches, and objects may remain associated with non-zero impact probabilities for extended periods---not because an impact is likely, but because their trajectories cannot be sufficiently constrained with available data. This defines a regime of \textit{follow-up–limited objects}, in which additional observations, such as those enabled by JWST, become essential to resolve otherwise persistent hazard assessments.

\begin{figure}[htb]
\hspace{-2mm}
\includegraphics[width=1\columnwidth]{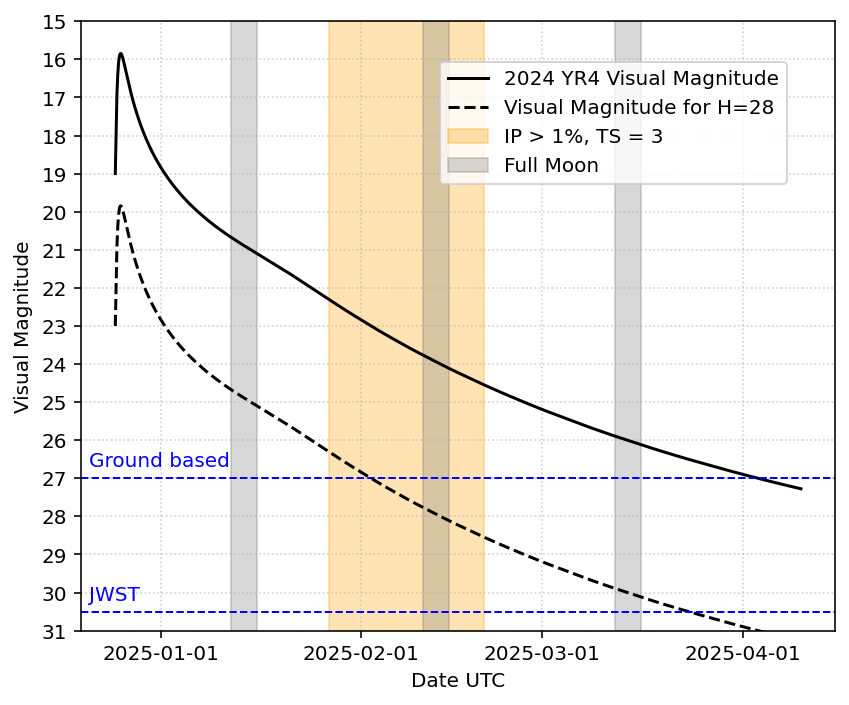}
\caption{\textbf{Brightness as a function of time for 2024 YR4 (solid line) and a smaller synthetic clone with the same orbit and an absolute magnitude $H=28$ (dashed line).}
The horizontal lines correspond to approximate brightness limits for ground-based telescopes ($V\sim27$) and JWST ($V\sim30.5$). The yellow shaded region marks the time interval during which 2024 YR4 had an Earth impact probability (IP) for 2032 greater than 1\%, corresponding to a Torino Scale (TS) rating of 3. While the real 2024 YR4 remained observable from the ground throughout this interval, the $\sim$10 m equivalent clone would have become inaccessible early in the high-IP phase, leaving its IP unresolved without additional observations from JWST.}

\label{fig:yr4_clone}
\end{figure}

As an example, we consider a synthetic clone of 2024 YR4 on the same orbit but with an absolute magnitude $H = 28$, which roughly corresponds to $\sim$10 m.
Fig.~\ref{fig:yr4_clone} shows the apparent $V$-magnitude of the real 2024 YR4 and the smaller clone.
The clone would have been sufficiently bright to be discovered in December 2024 during the close approach to Earth at 2 lunar distances.
However, it would have reached $V=27$, thus becoming too faint for ground-based telescopes, at the beginning of February, when the Earth impact probability was greater than 1\%.
Since the impact probability of 2024 YR4 dropped below 1\% on 2025 February 20, ground-based data would not have been sufficient to rule out the Earth impact for the smaller clone.
On the other hand, JWST would have enabled observations well into March and ruled out the possible impact.

Complementary approaches based on systematic precovery analysis can also help reduce or eliminate impact probabilities without requiring new observations. By extending the observational arc through the identification of pre-discovery detections, such methods can improve orbit solutions and reduce the number of cases requiring space-based follow-up (e.g., \citealt{Deen2026}).

Smaller objects are more numerous and therefore reach the Earth with higher frequency than larger ones. For example, an Earth impact of a 10-meter object happens roughly once every decade, on average \citep{Brown2002}.
Given the advent of surveys like the Vera Rubin Observatory \citep{Ivezic2019}, NEO Surveyor \citep{Mainzer2023}, NEOMIR \citep{Conversi2024}, and the Roman Space Telescope \citep{Holler2025}, and the resulting increases in NEO discovery rates at smaller sizes, we anticipate an increased need for JWST to support orbital refinement and impact hazard assessment. 


\section{Conclusion}
\label{sec:conclusion}

The JWST observations of 2024 YR4 demonstrate that large-aperture, highly sensitive space-based observatories can materially advance the timeline of hazard assessment for faint, decameter-scale objects. By extending the observational arc by eight months and reducing the uncertainty in the 2032 lunar encounter by more than an order of magnitude, the February 2026 measurements resolve the remaining impact probability approximately two years prior to the next feasible ground-based recovery. This gain in lead time is central to the operational value of such observations. Together with the Spring 2025 JWST/MIRI-based size measurements of 2024 YR4 \citep{Rivkin2025}, this result provides an operational realization of the role of JWST in planetary defense anticipated by \citet{Burdanov2025}, where the detection and characterization of decameter-scale main-belt asteroids highlighted its potential to monitor and constrain possible impactors.

In the context of increasing activity within the Earth--Moon system, such capabilities provide a means to assess the implications of impacts that, while not globally catastrophic, may carry operational relevance for orbital infrastructure and exploration planning. The sensitivity required to detect and track objects at flux levels of order $\sim 1$ nJy (i.e., $V\sim31$) places these observations well beyond the reach of other ground-based and space-based facilities (including survey-class missions).

Future large-aperture observatories such as the Habitable Worlds Observatory \citep{Dotson2025} may further extend this capability, enabling deeper sensitivity and broader access to faint-object populations.

Together, these results position 2024 YR4 not only as a resolved hazard assessment, but as a practical demonstration of how large-aperture space-based observations are likely to become central to planetary-defense efforts in the coming years, and beyond.


\section*{Acknowledgments}
This work is based on observations made with the NASA/ESA/CSA James Webb Space Telescope. The data were obtained from the Mikulski Archive for Space Telescopes at the Space Telescope Science Institute, which is operated by the Association of Universities for Research in Astronomy, Inc., under NASA contract NAS 5-03127 for JWST. These observations are associated with JWST DDT program \#9441, and we acknowledge the favorable review and prioritization of this project and key support from our program coordinator (A. Vick) and instrument reviewer (J. Stansberry).  Support for program number JWST-DD-9441 was provided through a grant from the STScI under NASA contract NAS5-26555. Part of this work was conducted at the Jet Propulsion Laboratory, California Institute of Technology, under a contract with NASA (80NM0018D0004). P.P. is supported by Praemium Academiae award by the Academy of Sciences of the Czech Republic, grant AP2401. D.J.T. is supported by NASA 80NSSC21K0807.

The data is available at MAST: \dataset[DOI: 10.17909/pec5-ke84]{https://doi.org/10.17909/pec5-ke84}.


\bibliography{bibliography}{}
\bibliographystyle{aasjournal}

\appendix
\label{Appendix}

\section{Details of the astrometric analyses}

This appendix provides the details of the astrometric analyses described in Section\,\ref{sec:astrometry}, followed by a discussion of the procedure followed to derive the astrometric timetags.

\subsection{Astrometric analysis by M. Micheli}

Astrometric measurements of 2024 YR4 at the two observed epochs were extracted using a hybrid approach designed specifically for this program, combining two different JWST pipeline data products.

The pixel coordinates of the target were measured on the standard level \texttt{2b i2d} images (the fully calibrated and rectified exposure-level product) from JWST's MAST archive, by fitting a radially symmetric Gaussian to the point spread function of the object. Centroiding uncertainties were estimated from the peak SNR of the detection and the fitted FWHM. These default \texttt{i2d} frames are ideal to analyze the target, because the various processing steps ensure an image that is clear from cosmic rays and other artifacts, but other objects such as field stars are significantly corrupted by the procedure.

The astrometric calibration for each image was instead performed on a custom-processed product generated by STScI staff, in which jump detection and saturation correction were both disabled. With this processing level, stellar trails remained visible and largely uncorrupted, enabling proper trail fitting of field stars from the Gaia DR3 catalog. The method assumed that the \texttt{i2d} products were fully linearized, and higher-order geometric distortions had already been corrected during the pixel remapping step that produced the \texttt{i2d} rectified frame, but allowed for a residual offset in the WCS solution in the form of a translation and possibly a small rotation with respect to the true sky plane. This offset, if present, was assumed to apply uniformly across all four NIRCam short-wavelength arrays. 
For each Gaia DR3 star visible in the image, the center of the corresponding trail was measured and compared to the position predicted by the WCS. Stars affected by saturation, lacking Gaia DR3 proper motion information, or truncated by detector edges were excluded. The collection of individual offsets was then averaged to yield a mean astrometric correction. A rotation term was also tested and found to be consistent with zero within uncertainties at the sub-arcminute level in all images: the WCS rotation was therefore held fixed throughout the analysis. For each exposure, the above procedure yielded a two-dimensional astrometric correction and its associated uncertainty, characterized by a 2D Gaussian ellipse with semi-axes and correlation coefficient derived from the distribution of individual stellar offsets. 

While this same general procedure was applied to both epochs, a few implementation details differed between the two nights. 
On the first epoch (2026 March 18), detector \texttt{nrcb1}, which contained the target, had an insufficient number of usable Gaia DR3 stars for a standalone solution, so the astrometric correction was derived from all four arrays jointly, and the shift was determined on the ensemble of all measured stars. Furthermore, the target's sky motion was slow enough that the entire exposure produced a nearly continuous trail of moderate length ($\sim$140 pixels), despite the presence of small $\sim$10 s gaps corresponding to the reset cycles between each integration. The trail fitting was therefore performed on the entire exposure-long trail.
On the second epoch (2026 March 26), \texttt{nrcb}1 alone contained enough usable reference stars for an independent chip-level solution, without considering the other three detectors. The object's motion was however significantly faster: this led to significantly longer full-exposure trails, but at the same time expanded the angular size of the inter-integration gap, making the trails corresponding to each integration clearly separated. It was therefore possible to trail-fit the central integration trail of each exposure in isolation (the 5th trail of the 9-integration readout mode used in the acquisition), allowing a more precise trail-fitting over a significantly smaller length.

The target's final astrometric position was finally obtained by applying the correction to the WCS-predicted coordinates of the object centroid, and the total astrometric uncertainty was computed as the convolution of the solution uncertainty ellipse with the circular centroiding uncertainty of the target. 

\subsection{Astrometric analysis by J. de Wit and A.Y. Burdanov}

Similarly to the previous analysis, this second astrometric pipeline adopts a hybrid approach, deriving the target’s pixel coordinates and the plate solution from two different data products. The step associated with the target centroiding is described in Section\,\ref{sec:photometry}; we focus here on the astrometric calibration.

A key distinction of this analysis is that it derives an astrometric solution using a single NIRCam detector, namely the detector hosting the target (nrcb1). This approach results in larger uncertainties due to (1) the smaller number of available reference stars (particularly in the first images of Visit~1) and (2) reduced leverage on the linear transformation from pixel coordinates to celestial coordinates. This is especially relevant given that the low level of residual distortion ($\sim$1 mas) supports the use of first-order plate solutions across all four NIRCam detectors. Despite these differences, this pipeline yields astrometric positions consistent with the other independent analyses within its $1\sigma$ uncertainties (Fig.\,\ref{fig:astrometry}, right panel).

Using the {\it NOJUMP\_NOSAT} data products, the pixel coordinates of stellar trails are derived through the following procedure. First, streaks are identified as contiguous groups of bright pixels. Second, a template is constructed to characterize typical streak properties (length, width, and orientation). Third, this template is used to generate an initial estimate of the streaks' centers. Fourth, each candidate streak is passed to a refinement routine that performs subpixel centroid estimation on a cropped region of the image.

\begin{figure*}[ht!]
\centering
\includegraphics[width=0.7\textwidth,trim=0cm 5.5cm 0cm 5cm, clip]{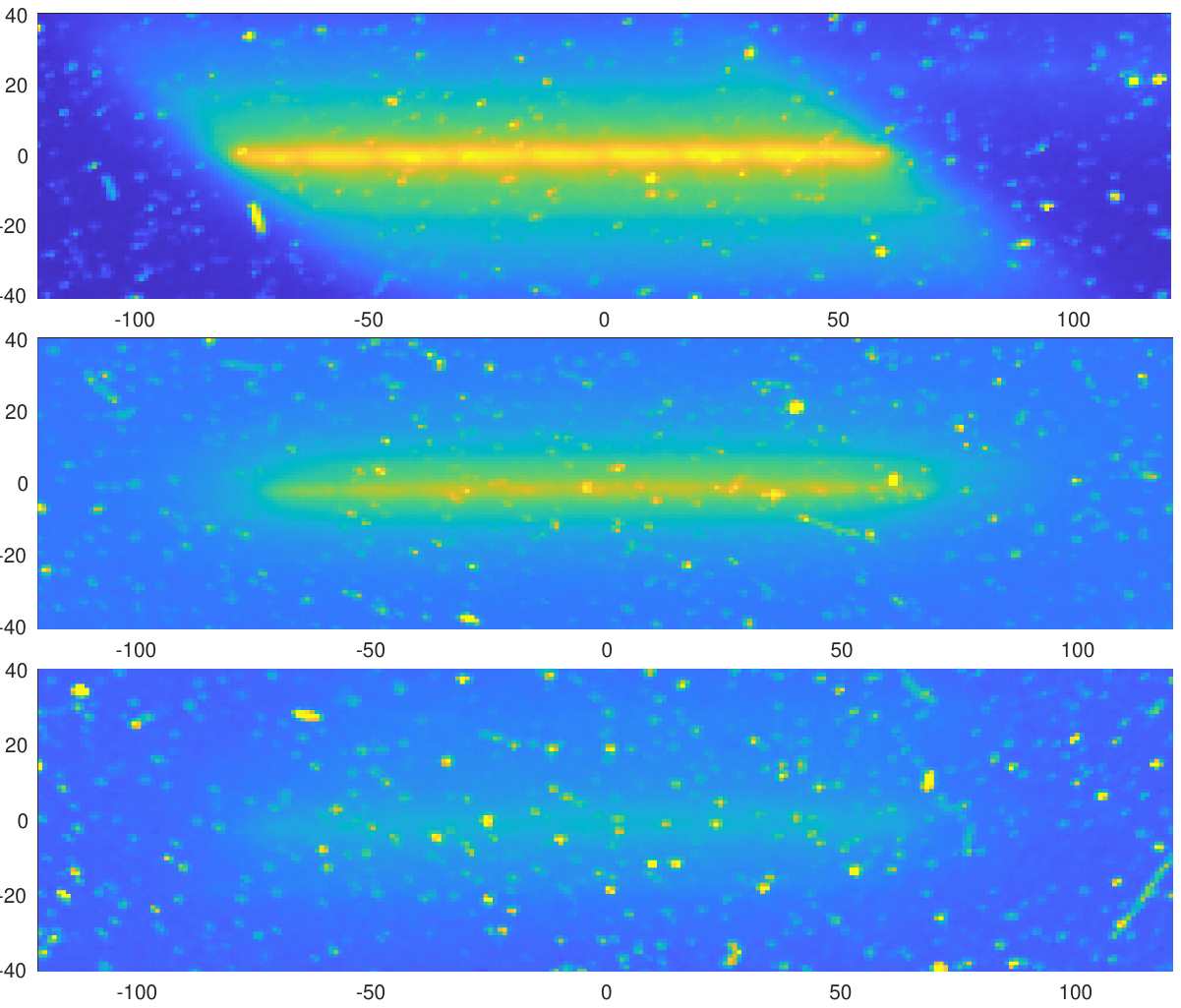}
\caption{\textbf{Examples of diffuse streaks in JWST/NIRCam imaging.}
Three representative streaks identified in the first image of Visit~1 are shown, contrasting with the well-defined, symmetric stellar trails used for astrometric calibration (see Fig.\,\ref{fig:astrometry}, left panel). These streaks exhibit extended or asymmetric morphologies, consistent with non-stellar sources such as background galaxies or solar system objects with intrinsic motion. These sources are not used in the astrometric solution presented in this work but illustrate the presence of additional populations detectable in these data, which may be explored in future analyses.
}
\label{fig:other_sources}
\end{figure*}

This procedure typically identifies 20 to 40 streaks per image, compared to 3 to 7 Gaia reference stars, depending on the image. Some of these streaks exhibit additional smearing, consistent with non-stellar sources (e.g., background galaxies or trans-Neptunian objects; see Fig.\,\ref{fig:other_sources}). These sources are not used in the astrometric solution and will be the subject of future analysis.

For each streak, the centroid is derived using a two-step approach. First, the streak is collapsed along its long axis to determine its weighted center in the cross-track direction. Second, the streak is collapsed along the short axis, and the centroid along the direction of motion is derived from the locations of the segment transitions associated with the nine groups in the readout pattern (see left panel of Fig.\,\ref{fig:astrometry}). The subpixel precision of these low-flux transitions enables a high-precision determination of the streak center, exceeding that obtained from direct template fitting alone. This framework also allows the centroiding of partially truncated streaks (e.g., near detector edges), even when only a subset of groups is available.

With the ensemble of measured pixel coordinates and the JWST WCS solution for each image, Gaia DR3 reference stars are identified and used for plate solving. Given the low level of residual distortion, first-order plate solutions are adopted. The robustness of the solution is assessed by generating ensembles of plate solutions using subsamples of reference stars (bootstrapping), with 10\,000 realizations per subsample in which stellar pixel positions are perturbed according to their uncertainties.

This approach enables the identification of unreliable reference stars, which can introduce systematic errors and are therefore excluded. A first example consists of Gaia sources lacking proper motion information. Another relates to highly saturated stars exhibit undetectable asymmetries, likely due to charge redistribution effects, leading to systematic offsets in their inferred centroids. These offsets are clearly revealed through comparisons of plate solutions derived from different subsets of reference stars and are consistent across images of a same visit, of order $\sim$250 mas in right ascension and $\sim$100 mas in declination in visit 1. These sources are therefore excluded from the final astrometric solution, although improved treatments of saturated or otherwise non-ideal reference stars may enable their reliable inclusion in future analyses, particularly in regimes with limited numbers of reference stars.

The ability of this pipeline to incorporate truncated streaks, combined with the use of first-order plate solutions, enables astrometric solutions to be derived from nrcb1 alone, albeit with larger uncertainties than when combining all four detectors (Fig.\,\ref{fig:astrometry}, right panel). Further refinements of this approach will be explored in future work.

\subsection{Astrometric analysis by D. J. Tholen}

The processing pipeline used by STScI to generate the final images for this third analysis
resulted in the presence of NaN (not-a-number) pixel values, which
caused issues for the astrometric software when such values were
present within either the object aperture or the background annulus
for the asteroid or reference stars. Prior to running the astrometric
analysis, a preprocessing step was therefore implemented to replace
NaN pixels with the average of the eight surrounding pixels.

The {\it JUMP} images optimized for the non-sidereal target were fit using a
two-dimensional Gaussian function with four free parameters: the x and y
pixel coordinates of the target centroid, along with the peak value and
width of the Gaussian. The background level was determined from the
median value in an annulus surrounding the target. The Gaussian fitting
was performed using pixels whose midpoints lie within a radius of 2.8
pixels centered on the asteroid. The inner and outer radii of the
background annulus were set to their default values of 2.5 and 5.0 times
the aperture radius (7 and 14 pixels, respectively). While centroid
positions are independent of pixel units, the associated uncertainties
depend on the assumption of Poisson statistics, requiring conversion
from Data Number units to photon counts. A region of the image free of
star trails and processing artifacts was used to estimate the noise and
derive an effective gain of 17031. The resulting centroiding uncertainty
is typically $\sim$0.2 pixel, corresponding to $\sim$6 mas.

The {\it NOJUMP\_NOSAT} images optimized for the reference star trails were fit using a
model consisting of a Gaussian profile in the cross-track direction and
a trapezoidal profile along the direction of motion. The six free
parameters are the x and y pixel coordinates of the trail center, the
peak value, and the length, width, and angle of the trail. 
This code was developed over two decades ago to measure ground-based images of
faint near-Earth asteroids and has been used for thousands of exposures with
excellent results.
Deviations
from perfect symmetry in the observed trails, introduced by the image
processing, are not explicitly captured by this model and may introduce
a small systematic bias, estimated to be at most $\sim$2 pixels
($\sim$60 mas). The small gaps in the trails as shown in Fig. 2 (left panel) raise the value of the
reduced chi-squared statistic, given that the model used to fit the trails does not
account for them, but do not affect the value of the trail center.

For the February 26 images, a sufficient number of Gaia DR3 reference
stars were available in each exposure to derive independent astrometric
solutions. For the February 18 images, two of the Gaia reference stars
are sufficiently bright to be affected by saturation and were therefore
excluded. As a result, only one exposure contained enough usable
reference stars for a fully independent solution. For the remaining
exposures, the available reference stars were used to determine the
astrometric zero point, while the image scale and orientation were
assumed to match those derived from the well-constrained exposure.

\subsection{Astrometric timetags}

The astrometry produced in this work is measured using stellar trails as reference sources. Since these stars are trailed due to the non-sidereal motion of the telescope, the resulting astrometry should be associated with the UTC time corresponding to the center of each trail, i.e., what would be the mid-exposure time in a traditional CCD.
However, NIRCam's readout process is significantly more complex, occurring on a timescale of 10.74 s, which is not negligible for astrometric purposes. The details of the readout process are therefore essential to determine the exact time of each astrometric measurement.

The first level of timing accuracy can be easily determined directly from the JWST data products. The level \texttt{2a rateints} FITS files, directly available on MAST, contain a table with the exact start, middle and end of each of the nine integrations composing each exposure, expressed directly in UTC (and in MJD). Since we have nine integrations, the middle of the fifth integration corresponds to the middle of the sequence and can be directly read from the ``int\_mid\_MJD\_UTC'' keyword.
Using this value as the exposure mid-time would be a crude approximation of the exact time, valid only under the assumption that the object falls exactly on the central row of the array.

The full readout cycle of NIRCam in full readout mode is 10.74 s long. Over each readout cycle, the camera rows are read sequentially from one end to the other. If the object is not located in the middle row, its signal would actually be read out earlier or later, introducing a time bias of up to $\pm$5.37 s at the extreme ends (first or last row).
To ensure the best possible precision for the astrometric timetags, we need to compensate for this offset by measuring the exact row where the asteroid falls and rescaling the time by the temporal offset between the readout of the central row and that of the row containing the target, taking into account the known readout direction of the specific chip where the target is located.
Furthermore, to achieve the best possible accuracy, the actual row needs to be measured on the original images, not on the linearized \texttt{i2d} processing level used for the astrometric extraction. These non-linearized frames can also be retrieved from MAST as the level \texttt{2b cal} data products. The asteroid's row can be located on these images and used to correct the timing that will then be associated with the final astrometric measurement.

This procedure was applied to all astrometric measurements presented in this work. To be safely conservative, additional timing uncertainties were reported in ADES format, assigning a 1.0 s possible timing bias (\texttt{uncTime} in ADES) and a 0.5 s possible random noise (\texttt{rmsTime} in ADES). These estimates are likely conservative but safely capture any possible unmodeled corrections that might remain in the process.

\end{document}